\documentclass{article}
\usepackage{epsfig}

\date{\today}

\begin{document}

 \title{ 
Spinning black strings
 in five dimensional\\ Einstein--Gauss-Bonnet gravity
} 

\author{{\large  Burkhard Kleihaus, Jutta Kunz,} 
{\large Eugen Radu} 
and {\large Bintoro Subagyo}  \\ 
 {\small  Institut f\"ur Physik, Universit\"at Oldenburg, Postfach 2503
D-26111 Oldenburg, Germany} }

\newcommand{\vphi}{\varphi}
\newcommand{\vepsilon}{\varepsilon}
\newcommand{\DS}{\displaystyle}
\newcommand{\pih}{\frac{\pi}{2}}
\newcommand{\sqdetg}{\sqrt{-g}}
\newcommand{\sqdetgi}{\frac{1}{\sqrt{-g}}}
\newcommand{\edil}{e^{2\kappa \phi}}
\newcommand{\bA}{\bar{A}}
\newcommand{\bF}{\bar{F}}
\newcommand{\bD}{\bar{D}}
\newcommand{\bx}{\bar{x}}
\newcommand{\beq}{\begin{equation}}
\newcommand{\eeq}{\end{equation}}
\newcommand{\beqs}{\begin{eqnarray}}
\newcommand{\eeqs}{\end{eqnarray}}
\newcommand{\reef}[1]{(\ref{#1})}
\newcommand{\tr}{\mbox{\rm tr}}
\newcommand{\ra}{\rightarrow} 
\newcommand{\be}{\begin{equation}}
\newcommand{\ee}{\end{equation}}
\newcommand{\bea}{\begin{eqnarray}}
\newcommand{\eea}{\end{eqnarray}}
\newcommand{\Ord}[2]{\mathcal O \left(#1\right)^{#2}}

\maketitle

 \begin{abstract}
We construct generalizations of the $D=5$ Kerr black string 
by including higher curvature corrections to the gravity action
in the form of the Gauss-Bonnet density.
 These uniform black strings satisfy a generalised Smarr relation and share the basic
properties of the
 Einstein gravity solutions.
However, they exist only up to a maximal value of the  Gauss-Bonnet coupling constant, 
which depends on the solutions' mass and angular momentum. 
\end{abstract}
 
\section{Introduction}
For a spacetime dimension $D>4$, the Einstein gravity presents a natural generalisation
-- the so called Lovelock theory, constructed 
from vielbein, the spin connection and their exterior derivatives without using the Hodge dual,
such that the field equations are second order  \cite{Lovelock:1971yv}, \cite{Mardones:1990qc}.
Following the Ricci scalar, the next order term in the Lovelock hierarchy is the Gauss-Bonnet (GB) one,
which contains quadratic powers of the curvature.
As discussed in the literature, this term appears as the first curvature stringy
correction to general relativity~\cite{1,Myers:1987yn}, when assuming
that the tension of a string is
large as compared to the energy scale of other variables.  
The action of the Einstein-Gauss-Bonnet (EGB) gravity is
\be
I = \frac{1}{16\pi G}\int d^D x \sqrt{-g}\left( R + \frac{\alpha}{4} L_{GB}  \right),
\label{action}
\ee
with
\be
L_{GB}=R^2 - 4 R_{\mu \nu}R^{\mu \nu} + R_{\mu \sigma \kappa \tau}R^{\mu \sigma \kappa \tau} ,
\ee
where $G$ is Newton's constant, $R$ is the Ricci scalar, $g$ is the determinant of the metric, $R_{\mu\nu}$ 
is the Ricci tensor, while $R_{\mu \sigma \kappa \tau}$ is the Riemann tensor.
The constant $\alpha$ in (\ref{action}) is the GB coefficient with dimension $(length)^2$ and is positive
in the string theory. 
The variation of the action (\ref{action}) with respect to the metric
tensor results in the EGB  equations 
\begin{eqnarray}
\label{eqs}
E_{\mu \nu }=R_{\mu \nu } -\frac{1}{2}Rg_{\mu \nu}
+\frac{\alpha}{4}H_{\mu \nu}=0~,
\end{eqnarray}
where
\begin{equation}
\label{Hmn}
H_{\mu \nu}=2(R_{\mu \sigma \kappa \tau }R_{\nu }^{\phantom{\nu}%
\sigma \kappa \tau }-2R_{\mu \rho \nu \sigma }R^{\rho \sigma }-2R_{\mu
\sigma }R_{\phantom{\sigma}\nu }^{\sigma }+RR_{\mu \nu })-\frac{1}{2}%
L_{GB}g_{\mu \nu }  ~,
\end{equation}
is the Lanczos (or the Gauss-Bonnet) tensor.
These equations contain no higher derivatives of the metric tensor than second order
and the model has proven to be free of ghost when expanding around flat space.

As expected, inclusion of a GB term in the gravity action leads to a variety of 
new features  (see \cite{Garraffo:2008hu}, \cite{Charmousis:2008kc}
for recent reviews of the  higher order gravity theories and their solutions). 
However, although the generalization of the spherically symmetric Schwarzschild-Tangherlini solution
in EGB theory has been known for quite a long time \cite{Deser}, \cite{Wheeler:1985nh},  
the issue of  solutions with compact extra dimensions is less explored.
Black string solutions, present for
$D\geq 5$ spacetime dimensions, are of particular interest, since they 
exhibit new features that have no analogue in the black hole case.

In the case of Einstein gravity,
the simplest solutions of this type 
are found by trivially extending to $D$ dimensions  
the vacuum solutions to Einstein equations in $D-1$ dimensions.
These then usually correspond to uniform black strings (UBSs)
with horizon topology $S^{D-3}\times S^1$.   
However,  this simple construction does
not generically  work in the presence of a GB term in the 
action \cite{kastor}.    
The only existing results  in the literature on UBSs with a GB term  concern the case of static configurations.
UBSs in five spacetime dimensions were discussed  in \cite{Kobayashi:2004hq}, 
as well as their $D>5$ $p-$brane generalizations
\cite{Sahabandu:2005ma}. 
The results there show  the existence of a number of new features in this case, for example
the occurrence
of a minimal value of the black strings' mass for a given 
GB parameter $\alpha$ (see also \cite{Suranyi:2008wc}). 
 The extension of the results in \cite{Kobayashi:2004hq}
 for all dimensions between five and ten
was given in  \cite{Brihaye:2010me}. 
 
The purpose of this work is to construct spinning generalizations of the 
known UBSs in EGB theory.
For simplicity, we shall restrict to the  case of five spacetime dimensions\footnote{The case $D=5$ is interesting from yet another
point of view, since the GB term appears there
in the low-energy effective action
for the compactification of the $M-$theory on a Calabi-Yau threefold  \cite{Antoniadis:1997eg}.}.
By solving numerically the field equations, we show that 
the $\alpha=0$ solution ($i.e.$ the Kerr black string) admits generalizations with a GB term
and discuss the new features which occur in this case.

\section{The model}

\subsection{Black strings in EGB theory: general formalism}  
 
In this work we are interested
in spinning solutions approaching asymptotically the 
four dimensional Minkowski-space times a circle, ${\cal M}_{4}\times S^1$.
The line element of this background is
\begin{eqnarray}
\label{KK-metric}
ds^2 =-dt^2 + dr^2  + dz^2   + r^2 d\Omega_{2}^2,
\end{eqnarray}
where the direction $z$ is periodic with period $L$,
 $r$ and $t$ are the radial and time coordinates, respectively, 
 while  $d\Omega^2_{2}=d\theta^2+\sin^2 \theta d\varphi^2$ is the unit metric on $S^{2}$.

 The physical quantities  of a spinning 
 configuration that can be measured asymptotically far away in the
transverse space are the mass  $M$, the tension ${\mathcal T}$ in the direction of the circle,
and the angular momentum $J$.
 Similar to Einstein gravity, these quantities are defined in terms of three constants
 $c_t,c_z$  and $c_\phi$ which enter the asymptotics of the metric functions
\begin{eqnarray}
\label{as2}
&g_{tt}\simeq -1+\frac{c_t}{r },~~~g_{zz}\simeq 1+\frac{c_z}{r },
~~~~g_{\varphi t}\simeq  \frac{c_\phi \sin^2 \theta}{r}.
\end{eqnarray} 
The mass, tension and angular momentum of a spinning black string solution are given by\footnote{For discussions
of the computation of charges in EGB theory without a cosmological
constant, see \cite{Deser:2002rt}.} 
\begin{eqnarray}
\label{2} 
M=\frac{V_{2}L}{16 \pi G}\left[2c_t-c_z \right],
~~{\mathcal T}=\frac{V_{2}}{16 \pi G}\left[c_t-2 c_z \right],
~~J=\frac{V_{2}L }{8 \pi G}c_\phi,
\end{eqnarray}
where $V_{2}=4\pi $ is the area of the unit $S^2$ sphere.

Similar to the static case,
one can also define a relative tension $n$ 
(also called the relative binding energy) 
\begin{eqnarray}
\label{3}
n=\frac{{\mathcal T} L}{M}=\frac{c_t-2c_z}{2c_t-c_z},
\end{eqnarray}
which measures how large the tension is relative to the mass. 
Uniform string solutions in vacuum Einstein gravity have $c_z=0$ and thus a 
relative tension $n=1/2$. 
However, $c_z$ does not vanish in the presence of a GB term, which leads to a relative tension $n\neq 1/2$
even in the static case \cite{Brihaye:2010me}.

The Hawking temperature of the solutions is given by
\begin{eqnarray}
\label{TH-gen} 
T_H=\frac{ \kappa}{2 \pi},
\end{eqnarray}  
with $\kappa$ the surface gravity.
The general results in \cite{Wald:1993nt} show that the entropy of a black object
($i.e.$ also of a black string) in EGB theory can be written
as an integral over the event horizon,
\begin{eqnarray}
\label{S-Noether}  
S=\frac{1}{4G}\int_{\Sigma_h} d^{3}x \sqrt{ h}(1+\frac{\alpha}{2}\tilde R),
\end{eqnarray} 
where $ h$ is the determinant of the induced metric on the horizon and $\tilde R$ is the event horizon curvature.

 The solutions should obey the first law of thermodynamics, which for spinning solutions
 contains an extra work term:
\be
dM = T_H dS + \mathcal T dL+\Omega_H dJ,
\label{firstlaw}
\ee
where $\Omega_H$ (the thermodynamic variable conjugate to $J$) is the event horizon velocity.

Interestingly, one can show that for solutions without a dependence on the extra-dimensions $z$,
the event horizon quantities $T_H,~S,~\Omega_H$ and the global charges  $M,~\mathcal T$
are related through the simple Smarr mass formula\footnote{This relation is obtained by starting from the Komar expressions,
and making use of the equations of motion 
and the expansion of the solutions at the horizon and at infinity.}
\begin{eqnarray}
\label{Smarr}
M-\mathcal T L= T_{\rm H} S+\Omega_{\rm H} J .
\end{eqnarray} 

An interesting feature of EGB gravity is the presence 
of two branches of static solutions, distinguished by their behaviour
for $\alpha \to 0$ \cite{Deser}. 
In this work we shall restrict our analysis to rotating UBSs 
 whose static limit
corresponds to the branch of static solutions 
with a well defined Einstein gravity limit. 
 
\subsection{The metric ansatz and boundary conditions}
Our solutions possess three Killing vectors $\partial_t$,
$\partial_\varphi$ and $\partial_z$
and are constructed within the following metric ansatz\footnote{
The choice in (\ref{metric}) of 
a conformal gauge for the $(r,\theta)$
sector of the metric, instead of the usual
 choice for Boyer-Lindquist 
coordinates,  leads to a more stable numerical scheme.
Also, 
for $\omega=0$, this line element describes static
UBSs in an 'isotropic' coordinate system  
(see the discussion in Section 4 of Ref. \cite{Kleihaus:2009dm}).
}
\begin{equation}
\label{metric} 
ds^2 = -fdt^2+\frac{m}{f}\left(dr^2+r^2 d\theta^2\right) 
       +\frac{l}{f} r^2 \sin^2\theta
          \left(d\varphi-\frac{\omega}{r}dt\right)^2+ p dz^2,
\end{equation}
where $f$, $m$, $l$, $p$ and $\omega$ are functions of $r$ and $\theta$, only.
The event horizon of these stationary black holes resides at a surface
of constant radial coordinate $r=r_{\rm H}$,
and is characterized by the condition $f(r_{\rm H})=0$.

At the horizon we impose the boundary conditions
\begin{eqnarray}
\label{BC-eh}
f\big|_{r=r_H}=m\big|_{r=r_H}=l\big|_{r=r_H}=0,~\omega\big|_{r=r_H}= \Omega_{\rm H} r_{\rm H},~\partial_r p \big|_{r=r_H} = 0. 
\end{eqnarray}
The boundary conditions at infinity,
\begin{eqnarray}
\label{BC-inf}
f\big|_{r=\infty}=m\big|_{r=\infty} =l\big|_{r=\infty}=p \big|_{r=\infty}=1,~\omega\big|_{r=\infty}=0, 
\end{eqnarray}
ensure that the solutions approach asymptotically the Kaluza-Klein background (\ref{KK-metric}).
Axial symmetry and regularity impose
the boundary conditions on the symmetry axis ($\theta=0$),
\begin{eqnarray}
\label{BC-axis}
\partial_\theta f\big|_{\theta=0} = \partial_\theta l\big|_{\theta=0} = 
\partial_\theta m\big|_{\theta=0} = \partial_\theta \omega\big|_{\theta=0} =
\partial_\theta p\big|_{\theta=0} = 0,
\end{eqnarray}
and, for solutions with parity reflection symmetry (the case in this work), agree with
the boundary conditions on the $\theta=\pi/2$-axis.
  The absence of conical singularities implies also $m=l$ at $\theta=0$.

 Expansion near the horizon 
in $\delta = (r-r_{\rm H})/r_{\rm H}$ 
yields to lowest order $f=\delta^2 f_2(\theta)$,
$m=\delta^2 m_2(\theta)$, $l=\delta^2 l_2(\theta)$, $\omega=\Omega_{\rm H} r_{\rm H}(1+\delta)$
and $p=p_0(\theta)+\delta^2 p_2(\theta)$.
The  metric of a spatial cross-section of the horizon reads
\begin{eqnarray}
 \label{eh-metric}
d\sigma^2=\frac{m_2(\theta)}{f_2(\theta)}r_H^2 d\theta^2+
\frac{l_2(\theta)}{f_2(\theta)} r_H^2 \sin^2\theta d\varphi^2
+p_0(\theta)dz^2,
\end{eqnarray}
the computation of the entropy from (\ref{S-Noether})
being straightforward, 
\begin{eqnarray}
 \label{S-new}
 \nonumber
&&
 S=\frac{\pi L}{2G}\int_0^\pi d \theta\bigg\{ r_H^2 \sin \theta \frac{\sqrt{l_2m_2p_0}}{f_2}
 +\frac{\alpha}{2} \sqrt{\frac{l_2p_2}{m_2}}
 \bigg [
 \sin \theta 
 \big(
 2
 -\frac{l_2''}{l_2}
 +\frac{m_2''}{m_2}
 -\frac{p_0''}{p_0}
 +\frac{3l_2'm2'}{4l_2m_2}
  +\frac{m_2'p_0'}{m_2p_0}
   \\
 &&
 {~~~~~~~~~~~~~~~~~~~~~~~~~~~~~~~~~~~~~} 
 -\frac{l_2'p_0'}{2l_2p_0}
 +\frac{l_2'^2}{2l_2^2} 
-\frac{3m_2'^2}{4m_2^2} 
 +\frac{p_0'^2}{2p_0^2} 
 \big)
 +\cos\theta 
 \big(
 \frac{3m_2'}{2m_2}
-\frac{2l_2'}{l_2}
-\frac{p_0'}{p_0}
 \big)
 \bigg]
 \bigg \},
\end{eqnarray}
(where a prime denotes $d/d\theta$).
Also, since $f_2$, $l_2$, $m_2$ and $p_0$ 
are strictly positive and finite for all values of $\theta$, and $z$ is a periodic coordinate, it is obvious that 
the solutions have an $S^2\times S^1$ event horizon topology.  The Hawking temperature $T_{\rm H}$ of the black strings
is
\begin{eqnarray}
 \label{TH}
 T_{\rm H}=\frac{1}{2 \pi r_{\rm H}}\frac{f_2(\theta)}{\sqrt{m_2(\theta)}}, 
\end{eqnarray}
the field equation  $E_\theta^r =0$ implying that the surface gravity 
is indeed constant on the horizon.
For completeness, we mention that 
the mass, tension and angular momentum are read from the asymptotic expansion (\ref{as2}),
with $g_{tt}=-f$, $g_{zz}=p$, $g_{\varphi t}=-l \omega r \sin^2\theta/f$.

\subsection{The equations and the Kerr black string}  
The scarcity of exact solutions is a generic feature of 
 EGB theory\footnote{In fact, for $D=5$, the only solution
 known in closed form corresponds to the generalization of the Schwarzschild-Tangherlini black hole
 found in \cite{Deser}, \cite{Wheeler:1985nh} (however, see also the spinning configurations with a negative cosmological
 constant in
 the Section V of \cite{Anabalon:2009kq}).}.
  For example, 
even in the static case, no closed form black string solution could be found  within a nonperturbative approach.
 Therefore we rely on numerical methods also on  constructing spinning UBSs.
   
The solutions in this work are found 
by using an approach originally proposed in \cite{Kleihaus:1996vi}
for $D=4$ solutions of Einstein gravity coupled with other matter fields,
which has been generalized in \cite{Kleihaus:2009dm}
to static solutions of the  $D=5$  EGB theory.

The equations for the functions  $\mathcal F_i=(f,l,m,\omega,p)$ 
we employ in the numerics,
are found by using a  suitable combination of the EGB equations,
$E_t^t =0,~E_r^r+E_\theta^\theta =0$,  $E_\varphi^\varphi=0$, $E_z^z=0$ 
and  $E_{\varphi}^{t} =0$,
 which diagonalizes the Einstein tensor $w.r.t.$ $\nabla^2 \mathcal F_i$ 
 (where $\nabla^2=\partial_{rr}+\frac{1}{r}\partial_{r}+\frac{1}{r^2}\partial_{\theta\theta}$). 
 The remaining equations $E_\theta^r =0,~E_r^r-E_\theta^\theta  =0$
yield two constraints. 
Following \cite{Wiseman:2002zc}, one can show that
the identities $\nabla_\mu E^{\mu r} =0$ and $\nabla_\mu E^{\mu \theta}=0$, 
imply the Cauchy-Riemann relations
\begin{eqnarray}
\partial_{\bar r} {\cal P}_2  +
\partial_\theta {\cal P}_1  
= 0 ,~~
 \partial_{\bar r} {\cal P}_1  
-\partial_{\theta} {\cal P}_2
~= 0 ,~~~{~~~} 
\end{eqnarray}
with ${\cal P}_1=\sqrt{-g} E^r_\theta$, ${\cal P}_2=\sqrt{-g}(E^r_r-E^\theta_\theta)/2$
and $d\bar r=\frac{dr}{r}$.
Therefore the weighted constraints still satisfy Laplace equations, and the constraints 
are fulfilled, when one of them is satisfied on the boundary and the other 
at a single point
\cite{Wiseman:2002zc}.

The resulting set of five second order  coupled non-linear 
 partial differential equations\footnote{Due to the
GB contribution, these equations are much more complicated than in the case of Einstein
gravity (with more than 100 terms each equation). Then we shall not present them here.} 
for the functions $\mathcal F_i$ is solved numerically,
subject to the boundary conditions (\ref{BC-eh})-(\ref{BC-axis}), employing a compactified  
coordinate\footnote{Therefore we restrict the numerical
integration to the region outside the horizon,
$r\geq r_H$.}  $x = 1 - r_{\rm H}/r$,
which  leads to
a rectangular shape for the domain of integration, $0\leq x\leq 1,~0\leq \theta \leq \pi/2$.
The numerical calculations are based on the Newton-Raphson method
and are performed with help of the program FIDISOL/CADSOL \cite{schoen},
which provides also an error estimate for each unknown function.
For the solutions in this work,
the typical  numerical error 
for the functions is estimated to be lower than $10^{-3}$. 
The Smarr relation (\ref{Smarr}) provides a further test of the numerical accuracy. 

In this approach, one provides the input parameters 
$(\alpha;~r_H,~\Omega_H)\geq 0$.
The quantities of interest are computed from the numerical output
(for example, the mass $M$, tension $\mathcal T$ and angular momentum $J$
are extracted from the  asymptotic expressions (\ref{as2})).

 The equations satisfied by the metric functions are invariant under the following rescaling:
\be
\label{scaling}
\alpha \rightarrow \lambda^2 \alpha,\ r \rightarrow \lambda r \ .
\ee
Then a dimensionless relevant parameter can   be defined according to
$\beta = \alpha/\lambda^2 $,
where $\lambda$ is some length scale.
Following  \cite{Sahabandu:2005ma}, \cite{Brihaye:2010me}, 
we have found it convenient to choose $\lambda$ as the horizon radius $r_H$ 
of the black string and thus to define
\be
\label{beta_def}
\beta \equiv \frac{\alpha}{r_H^2}.
\ee
Also, for UBS solutions, the period $L$ of the 
$z$-direction
is an arbitrary positive constant and plays no role in 
 our results.
 Then, to simplify the relations,  we set $L=G=1$
 in all results below
  ($i.e.$ one considers the values of $M,S$ and $J$ per unit length of the extra-dimension).

The Kerr black string is recovered for $\alpha=0$ and has $g_{zz}=p(r,\theta)=1$, the expression 
of the other metric functions (for the metric ansatz (\ref{metric})) being
  \begin{eqnarray}
  \label{Kerr}
 f=\left(1-\frac{r_H^2}{r^2}\right)^2\frac{F_1}{F_2},~~
 l=\left(1-\frac{r_H^2}{r^2}\right)^2,
 ~~
 m=\left(1-\frac{r_H^2}{r^2}\right)^2\frac{F_1^2}{F_2},~~
 \omega=\frac{2M\sqrt{M^2-4r_H^2}}{r^2}\frac{(1+\frac{M}{r}+\frac{r_H^2}{r^2})}{F_2},
 \end{eqnarray}
 where
\begin{eqnarray}
\nonumber
&&
F_1=\frac{2M^2}{r^2}+\left(1-\frac{r_H^2}{r^2}\right)^2+\frac{2M}{r}\left(1+\frac{r_H^2}{r^2}\right)-\frac{M^2-4r_H^2}{r^2}\sin^2\theta,~~
\\
&&
\nonumber
F_2=\left(\frac{2M^2}{r^2}+\left(1-\frac{r_H^2}{r^2}\right)^2+\frac{2M}{r}\left(1+\frac{r_H^2}{r^2}\right)\right)^2
-\left(1-\frac{r_H^2}{r^2}\right)^2\frac{M^2-4r_H^2}{r^2}\sin^2\theta.
\end{eqnarray}
The value of the horizon velocity (which enters the boundary conditions at $r=r_H$)
is expressed in terms of mass and event horizon radius as  $\Omega_H=\frac{\sqrt{M^2 -4r_H^2}}{2M^2+4Mr_H }$, with $M\geq 2 r_H$.
The   entropy, angular momentum and the Hawking temperature of the Kerr UBS are 
given by $S=2 \pi M(M+2 r_H)$, $J=M\sqrt{M^2-4 r_H^2}$
and $T_H=\frac{1}{4\pi M}\frac{1}{1+\frac{M}{2r_H}}$, respectively.

 In studying the solutions' properties,
 it is convenient to work with 'reduced' dimensionless
 quantities as follows:
  \begin{eqnarray}
~~t_H=8 \pi T_H M,~~
j=\frac{J}{M^2},~~a_H=\frac{1}{16 \pi}\frac{A_H}{M^2},
~~s=\frac{1}{4\pi}\frac{S}{M^2},
 \end{eqnarray}
 the length scale being fixed in this case by the mass of the solutions.
 For the solutions without a GB term, one finds from (\ref{Kerr}) the simple relations 
  $a_H=\frac{1}{2}(1+\sqrt{1-j^2}),~~t_H=\frac{2\sqrt{1-j^2}}{1+\sqrt{1-j^2}},
 ~~s=\frac{1}{2}(1+\sqrt{1-j^2}).$
 
 Alternatively, following \cite{Kobayashi:2004hq},
one can scale all quantities with respect to $\alpha$, 
taking into account the corresponding dimensions
($e.g.$ $S  \sim (length)^{2}$, $\Omega_H \sim (length)^{-1}$ etc).

\section{The results}  
\subsection{The static black strings}  

Before discussing the spinning UBSs, let us briefly review the situation in the static case.
For the metric ansatz (\ref{metric}),
these solutions are found in the limit $\omega=0$ and have $l=m$, with $f,m,p$ functions of $r$ only.
The static UBSs were studied within a nonperturbative approach in \cite{Kobayashi:2004hq},
\cite{Brihaye:2010me} and \cite{Kleihaus:2009dm}.
The numerical results there show the existence, for  a given value of $r_H$, 
of a maximal value of $\alpha$, with\footnote{Note that the results 
in \cite{Kobayashi:2004hq} and \cite{Brihaye:2010me}
were found for a Schwarzschild-like coordinate system.
The value of the event horizon radius in that case differs from $r_H$
for the 'isotropic' line-element (\ref{metric}), 
which translates into a different maximal value of the parameter $\beta$.
}  
$\beta^{(max)}=\alpha^{(max)}/r_H^2\simeq 5.8$.

Since for $\alpha>0$ there is a finite minimal value of the horizon radius,
this entails the existence of a minimal value of
the mass\footnote{Note that this feature is absent for black strings in more than five dimensions  \cite{Brihaye:2010me}.} 
for a given GB coupling constant $\alpha$, 
a property which is inherited  by the static EGB black rings
approaching asymptotically the ${\cal M}_5$ background \cite{Kleihaus:2009dm}.
This strongly constrasts with the picture found for EGB black holes with an
$S^3$ topology of the horizon, where $\alpha$ takes arbitrary values.

Interestingly, the static UBSs admit an analytic expression as a power series in $\alpha$ around the Einstein gravity solution.
The perturbative solution reads
\begin{eqnarray}
\label{rel1}
f(r)=f_0(r)\bigg(1+\sum_{k=1}^{\infty}\alpha^k f_k(r) \bigg),~~m(r)=m_0(r)\bigg(1+\sum_{k=1}^{\infty}\alpha^k m_k(r) \bigg),~~
p(r)= 1+\sum_{k=1}^{\infty}\alpha^k p_k(r),
\end{eqnarray} 
with
$f_0=\big(\frac{1-\frac{r_H}{r}}{1+\frac{r_H}{r}}\big)^2,~m_0=(1-\frac{r_H^2}{r^2})^2,$
the metric functions for the UBS solution in Einstein gravity.
One finds $e.g.$ for the first order solution
\begin{eqnarray}
\label{rel2-0}
 f_1(r)=-\frac{1}{2r^2(1+\frac{r_H}{r})^6}
\bigg(
(1-\frac{r_H}{r})^2
+\frac{44r_H}{9r}
+\frac{r}{6 r_H}(1-\frac{r_H^2}{r^2})^2 
\bigg)~~{\rm and}~~m_1(r)=-p_1(r)=2 f_1(r).
\end{eqnarray} 
The expression of the solution becomes very  complicated for higher values of $k$ and we shall not give it here.

Once the (perturbative) solution is known, it is straightforward to extract the relevant global quantities.
One finds in this way that, to second order in $\alpha$, the following relations hold
(with $M_0= 2r_H, ~{\cal T}_0=r_H,~S_0=16 \pi r_H^2$ and $T_H^0=\frac{1}{16\pi r_H}$
  the mass, tension, entropy and Hawking temperature in Einstein gravity):
\begin{eqnarray}
\label{rel2}
& M=M_0\big (1+\frac{1153 \beta^2}{7096320}  \big),~{\cal T}={\cal T}_0\big (1-\frac{\beta}{16}-\frac{1129 \beta^2}{1182720}\big),
~S=S_0 \big (1+\frac{\beta}{16}+\frac{14557 \beta^2}{7096320} \big),~T_H=T_H^0\big (1+\frac{1153 \beta^2}{7096320}\big),~{~~}
\end{eqnarray} 
which provides a reasonable approximation for the (numerical) nonperturbative results.
Inclusion of higher order terms in (\ref{rel2}) does not change this pattern:
the mass, entropy and temperature increase with $\beta$, while
the tension  decreases.
At the same time, the linear terms in $\beta$  
are absent in the expressions of  $M$ and $T_H$.
As a result, the mass and the temperature of a static EGB black string with a given event horizon radius $r_H$ do not change
significantly with the GB parameter $\alpha$.

\subsection{Spinning solutions}  
In principle, the slowly rotating UBS can be constructed in closed form, by taking the
perturbative solution (\ref{rel1})
 for the static background.
For example, to lowest order in $\alpha$, the expression of the metric function
associated with rotation is
$\omega =\frac{a(1+\alpha j_1)}{r^2(1+\frac{r_H}{r})^6}$,
with $a$ the small rotation parameter.

This is linear in the perturbation parameter $a$, while the
other functions remain unchanged to this order in $a$.
However, this  approach has some obvious limitations and we shall not pursue it here. 

The nonperturbative solutions are found by directly solving the EGB equations for the functions ${\cal F}_i$
without any approximation.
As expected, we have found numerical evidence that the spinning Einstein gravity solution (\ref{Kerr}) 
also admits generalizations
with a GB term.
These solutions are found by starting with the Kerr metric (with given $r_H,\Omega_H$)
 as the initial guess, and slowly increasing the value of $\alpha$.
 The iterations converge, and repeating the procedure one obtains in this way
solutions with large $\alpha$.

For all solutions we have found,
the metric functions $\mathcal F_i$ and their first and second derivatives with respect to $r$ and $\theta$ have 
smooth profiles, which leads to finite curvature invariants on the full domain of integration, in particular
on the event horizon.
The shape of the functions $f,l,m$ and $\omega$ is similar to the $\alpha=0$ case, the maximal deviation from the
Einstein gravity profiles being around the horizon.
As expected, $\alpha\neq 0$ leads to a metric function $g_{zz}\neq 1$, which in the rotating case,
possesses a nontrivial angular dependence, see Figure 1 (left).

\setlength{\unitlength}{1cm}
\begin{picture}(8,6) 
\put(-0.5,0.0){\epsfig{file=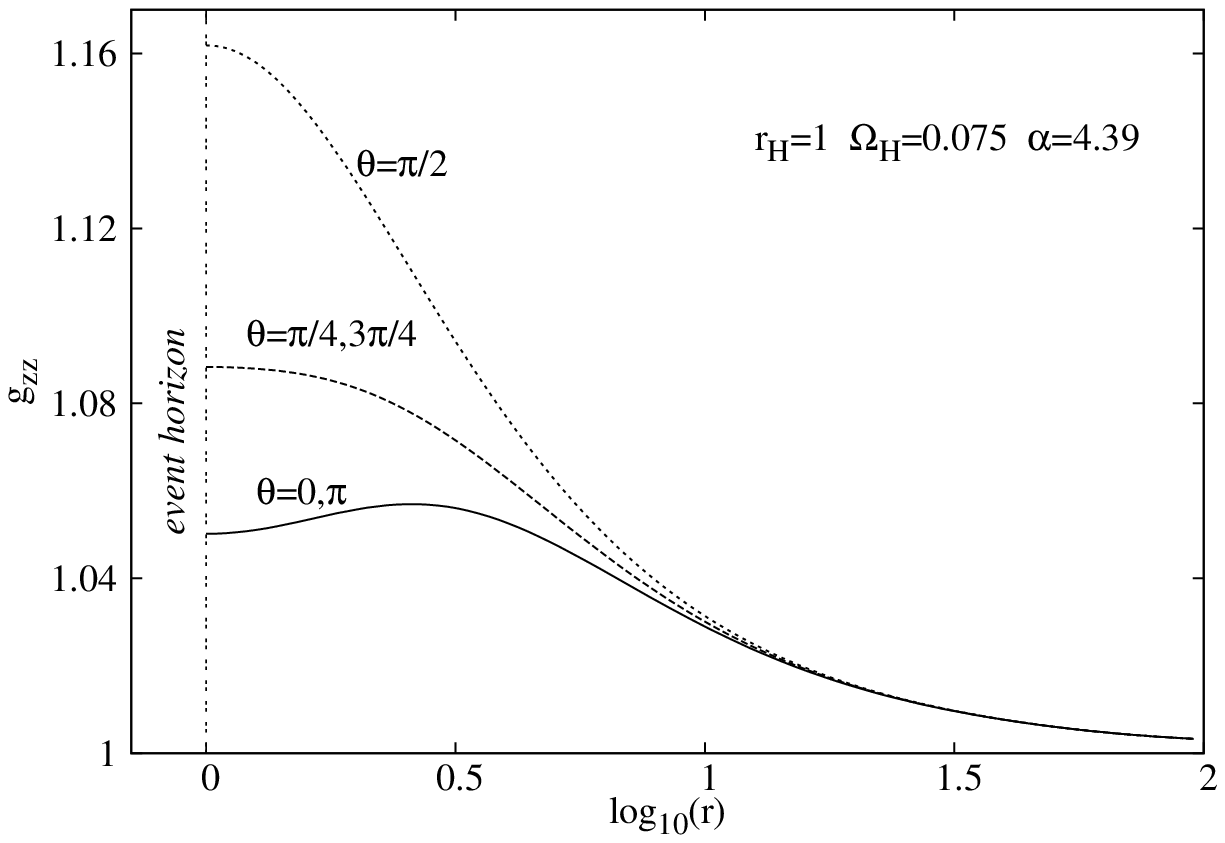,width=8cm}}
\put(8,0.0){\epsfig{file=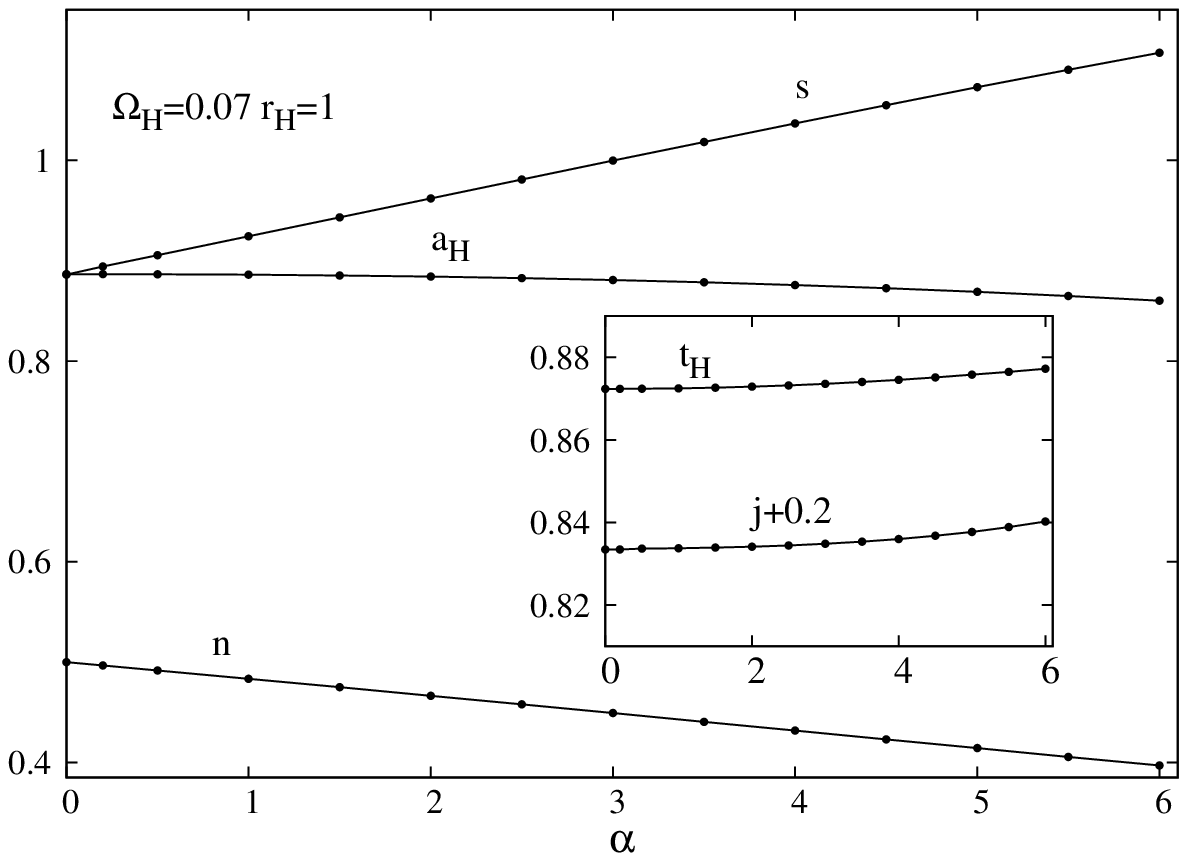,width=8cm}}
\end{picture}
\\
\\
{\small {\bf Figure 1.} {\it Left}.
The metric function $g_{zz}$ is shown as a function of $r$ and several values of $\theta$
for a typical spinning black string in EGB theory.
{\it Right}.
 A number of reduced parameters are shown as a function of $\alpha$
for a set of black strings with fixed event horizon velocity $\Omega_H$ and fixed event horizon radius $r_H$.
Here and in Figures 2, 3, the dots represent the data points while
the curves are obtained by spline-interpolation.
Also, for all data displayed in this work we set $L=G=1$.
   }
\vspace{0.5cm}

The general pattern is, however, quite complicated, and depends on the value of
the parameter $\alpha$.
As one can see in Figure 1 (right), for given $(r_H,\Omega_H)$, the relative
tension $n$ and scaled horizon area $a_H$  decrease with $\alpha$,
while the scaled temperature $t_H$, entropy $s$ and angular momentum $j$ increase.
The picture there seems to be generic and has been recovered for other
values of $(r_H,\Omega_H)$.

For given values of the horizon input data ($r_H,\Omega_H$), 
we have noticed the existence of a maximal value of 
the GB parameter $\alpha$.
This translates into a maximal value of the ratio $\alpha/M^2$
for a given value of the reduced angular momentum $j=J/M^2$.
For example, for $j=0$,
one finds $\alpha/M^2 <1.31$.
However, in the spinning case, it is rather difficult to provide such estimates,
since both $M$ and $J$ are output parameters and cannot easily be kept fixed.

As  $\beta^{(max)}=\alpha^{(max)}/r_H^2$
is approached,
the numerical process fails to converge, although no singular behaviour
is found there. The technical reason which causes the solutions to
cease to exist at $\beta^{(max)}$ is similar to the static case (see $e.g.$ the discussion in \cite{Kleihaus:2009dm}), and 
can be seen in the horizon expansion of the metric functions.
One finds that, for given $(r_H,\Omega_H)$, the roots of a quadratic  equation in the horizon parameters
$p_0,~m_2,~f_2$ cease to be real at $\beta^{(max)}$. 
We mention that the same behaviour has been noticed for other non-spherically symmetric solutions
with a GB term in the action, see \cite{Kleihaus:2009dm}, \cite{Kleihaus:2011tg}.

However, for the allowed range of $\beta=\alpha/r_H^2$, the overall picture is rather similar to the case of 
Einstein gravity, any static black string admitting
rotating generalizations. 
Here it is instructive to keep fixed the 
 parameter $\beta$
and to study the effects of an increasing event horizon velocity
on the properties  of UBSs (these solutions are found by starting with the static solutions in \cite{Kobayashi:2004hq} 
(written, however, in 
the 'isotropic' coordinate system (\ref{metric})) 
and slowly increasing the event horizon velocity $\Omega_H$).
Some numerical results in this case are shown in Figures 2, 3.

When increasing $\Omega_H$ from zero, while keeping $(r_H,\alpha)$ fixed, a branch of spinning UBS
solutions forms, the lower branch.
It extends up
to a maximal value of $\Omega_H$,
where an upper branch emerges and bends backwards towards
$\Omega_H=0$.
The maximal value of $\Omega_H$ depends on $(r_H,\alpha)$, with 
$\Omega_H^{(max)}=\frac{1}{2r_H}\frac{\sqrt{\frac{2}{1+\sqrt{5}}}}{3+\sqrt{5}} $
for $\alpha=0$.
Our results show that the value of $\Omega_H^{(max)}$ slowly decreases with 
$\beta$ by a simple scaling.
Along both branches, the mass,  tension,  entropy 
and angular momentum continuously increase\footnote{We emphasize that the existence of two branches of
solutions in terms of $\Omega_H$ for given $r_H$
is a result of 
using an 'isotropic' coordinate system in (\ref{metric}), and it occurs already for the
Kerr UBS.}.
Interestingly, the relative tension $n$ increases also with the angular momentum (see Figure 2 (right)), and 
appears to approach asymptotically 

\setlength{\unitlength}{1cm}
\begin{picture}(8,6)
\put(-0.5,0.0){\epsfig{file=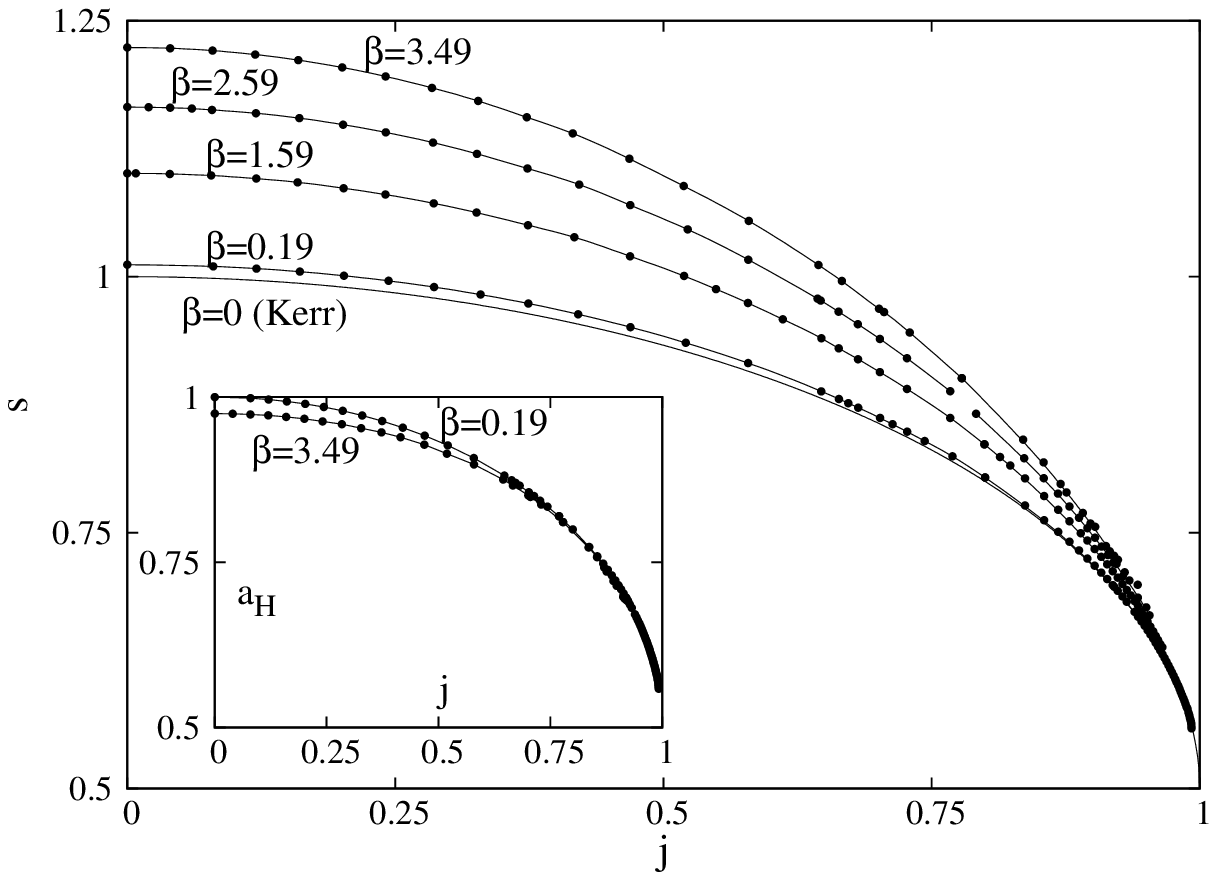,width=8cm}}
\put(8,0.0){\epsfig{file=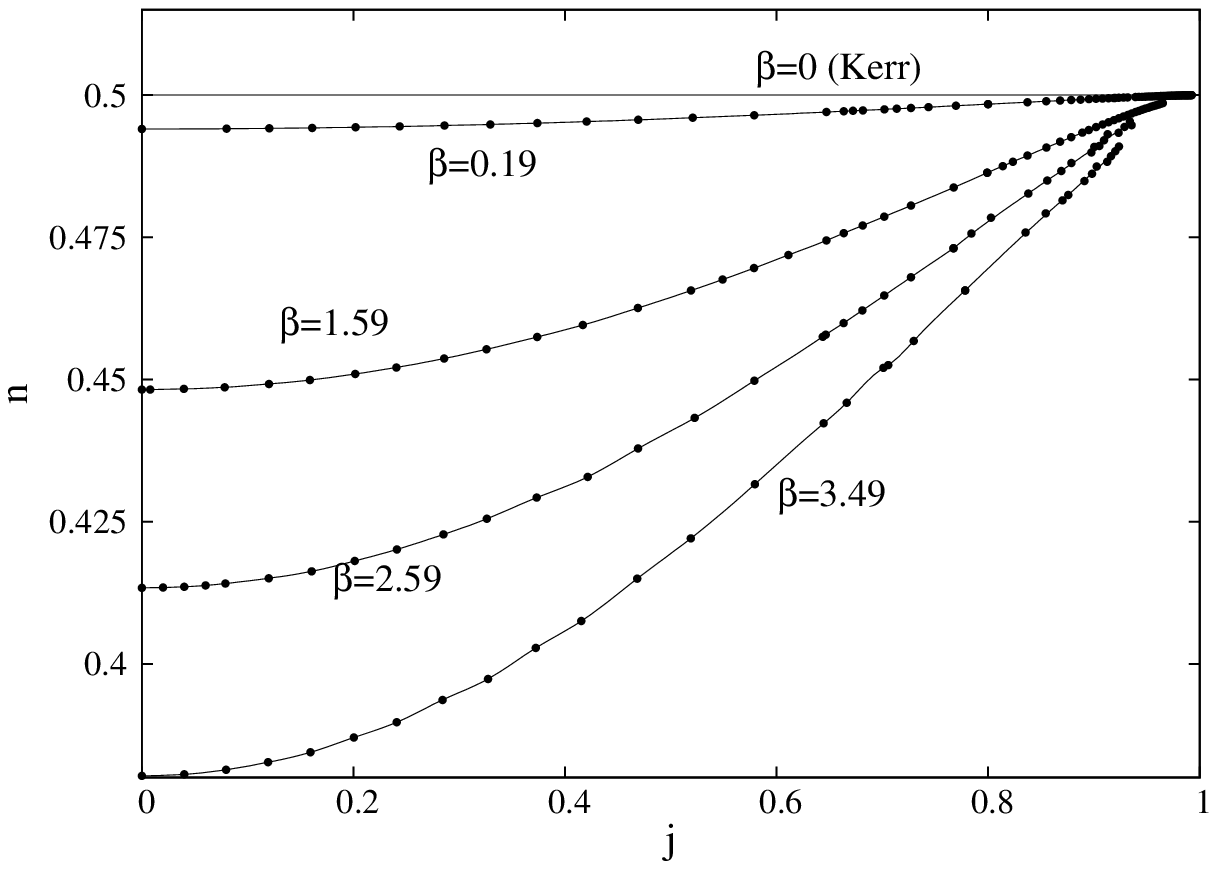,width=8cm}}
\end{picture}
\\
\\
{\small {\bf Figure 2.} The reduced entropy  $s=\frac{1}{4\pi}\frac{S}{M^2}$ and area $a_H=\frac{1}{16 \pi}\frac{A_H}{M^2}$ 
(left) and 
the relative tension  $n={{\mathcal T} L}/{M}$ (right) are plotted $vs.$ the reduced angular momentum $j=J/M^2$ for
  several values of  the parameter $\beta=\alpha/r_H^2$. 
   }
\vspace{0.5cm}
 
\setlength{\unitlength}{1cm}
\begin{picture}(8,6)
\put(-0.5,0.0){\epsfig{file=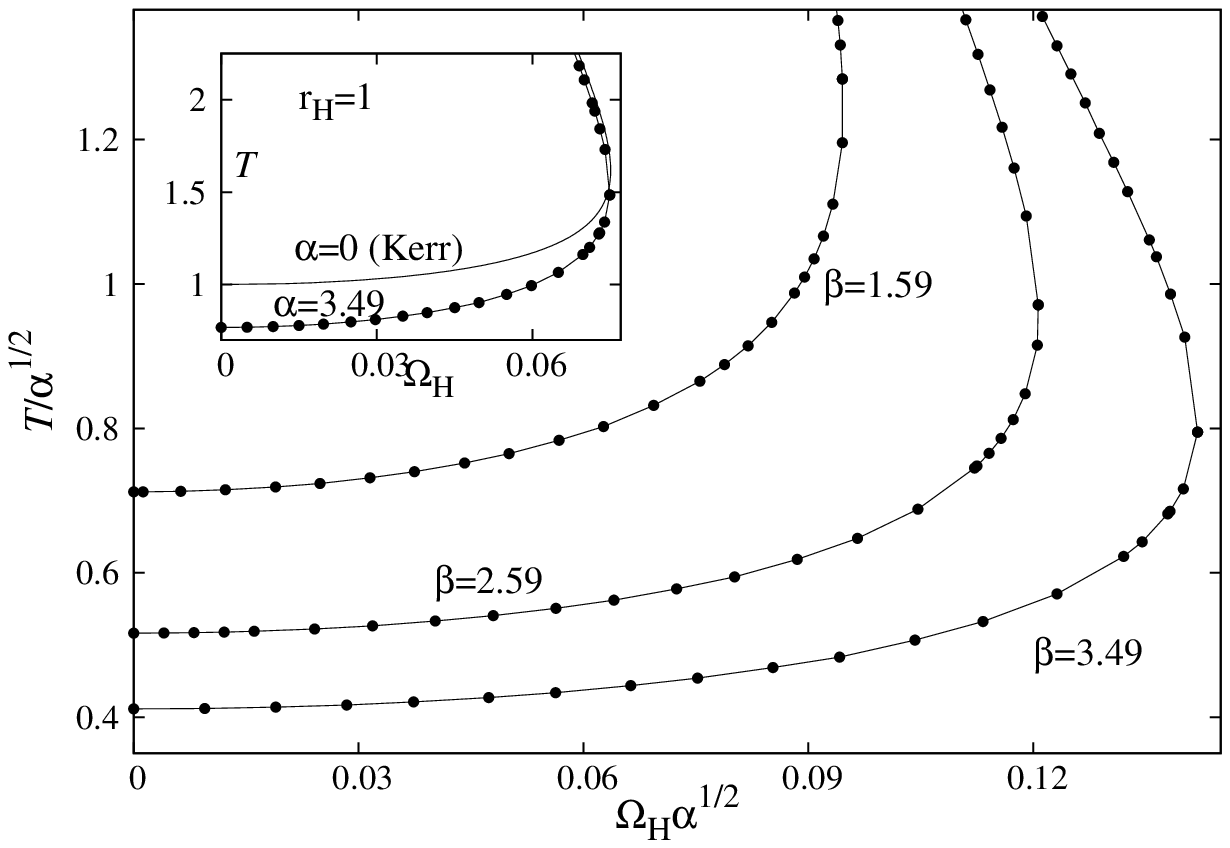,width=8cm}}
\put(8,0.0){\epsfig{file=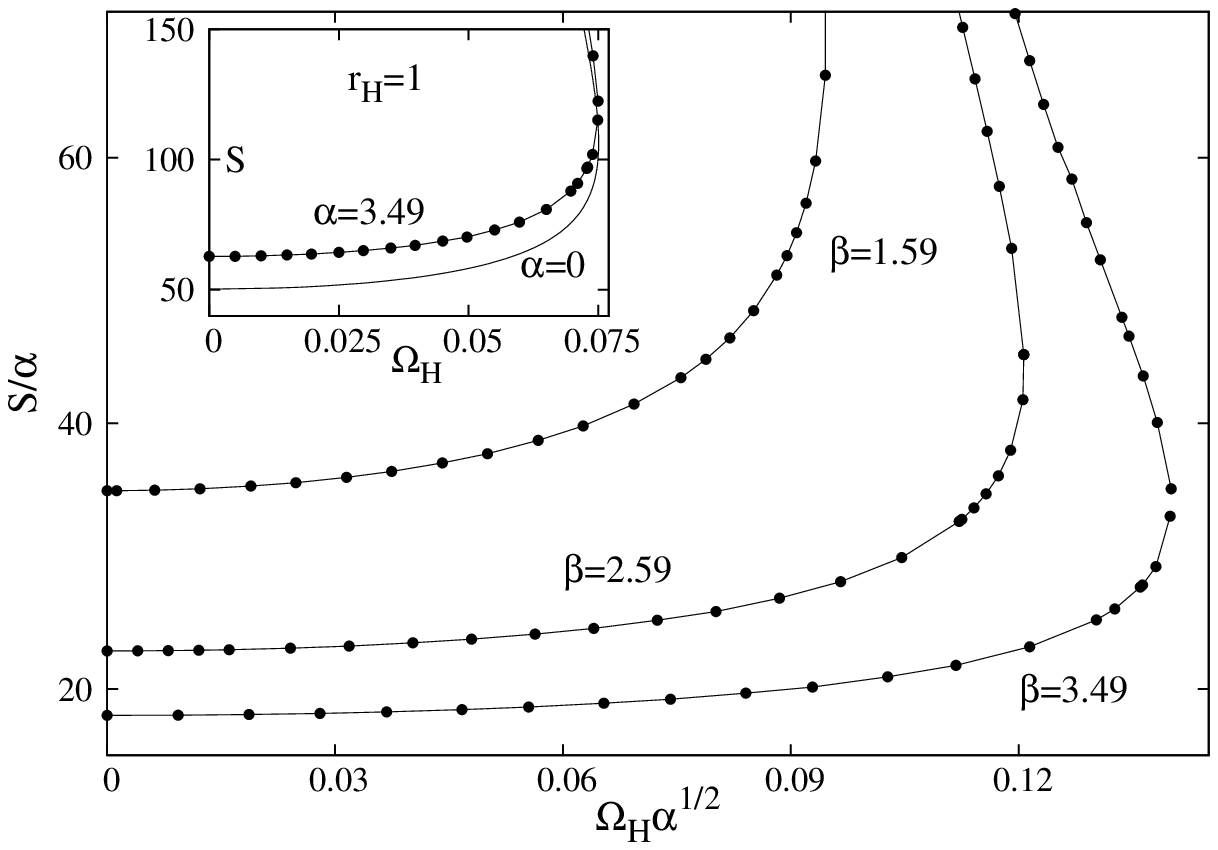,width=8cm}}
\end{picture}
\\
\\
{\small {\bf Figure 3.} The tension  and the
entropy are plotted $vs.$ the angular momentum velocity for
  several values of the parameter $\beta=\alpha/r_H^2$, 
the quantities being given in units of the Gauss-Bonnet constant $\alpha$.
 The insets show a comparison between the picture for  Einstein and Einstein-Gauss-Bonnet
 gravity solutions 
 with the same value of the horizon radius (unscaled quantities).
   }
 \vspace{0.5cm}
 \\
 the Kerr value $n=1/2$ for solutions
with $\Omega_H \to 0$ on the upper branch ($i.e.$ $c_z/c_t\to 0$ in that limit).

Also, we have noticed that, for a given $\beta$, the mass and Hawking temperature have only a small deviation 
from the corresponding values in the Einstein gravity case,   
while the angular momentum, entropy and tension change significantly.
We expect that the explanation of this behaviour would be similar to that found in the static case,
namely that no terms linear in $\alpha$ will enter the expressions of $M$ and $T_H$ 
in the rotating generalization of the perturbative result (\ref{rel2}).

\subsection{The issue of extremal black strings}  

For all considered values of $\beta=\alpha/r_H^2$, the numerical iteration fails to converge for 
solutions on the second branch with small values of $\Omega_H$.
In that limit, the Hawking temperature takes very small values, which suggests that
the limit  $\Omega_H \to 0$ corresponds to an extremal configuration.
For example, the family of solutions with $\alpha=0$ ends at the extremal Kerr UBS,
which precisely saturates the Kerr bound for the scaled angular momentum.

A study of the extremal UBSs  would require a 
different metric ansatz
than (\ref{metric}) and is beyond the purposes of this work.  
However, we argue that, different from the $\alpha=0$ extremal Kerr solution,
the extremal UBSs with GB corrections are likely to not represent regular configurations.  
This is supported by our results when attempting to 
construct the corresponding 
near-horizon geometries with an isometry group $SO(2,1)\times U(1)\times U(1)$.

There, following the usual ansatz in the literature (see $e.g.$ \cite{Astefanesei:2006dd})
 we consider the  line element
\begin{eqnarray}
\label{metric-eh}
 ds^2=v_1(\theta) \left ( -\rho^2 dt^2+\frac{d \rho^2}{\rho^2}+\bar \beta^2 d\theta^2 \right)
 + \bar \beta^2 v_2(\theta) \left(d\phi+ K \rho dt \right)^2+v_3(\theta)dz^2,
\end{eqnarray}
where $0\leq \rho<\infty$, $0\leq \theta \leq \pi $,  and $\bar\beta,~K$
 are real parameters. The above line element
 describes the
neighbourhood of the event horizon of an extremal UBS (and will 
be an attractor for the full bulk solutions).

Within this ansatz, the EGB equations (\ref{eqs}) result 
in a set of coupled nonlinear
ordinary differential equations.
For $\alpha=0$, the Einstein gravity solution is recovered, with \cite{Horowitz}
 \begin{eqnarray}
\label{solE}
 K=\bar \beta=1,~~v_1=\frac{J}{32 \pi}(3+2\cos 2\theta),~~v_2=\frac{J}{4 \pi}\frac{ \sin^2\theta}{(1+\cos^2 \theta)},~v_3=1,
 \end{eqnarray}
and $J>0$ an integration constant.

Unfortunately, no closed form solution could be found  in the presence of a GB term.
Therefore, as a first step, 
we have considered a perturbative
 solution in $\alpha$ of the EGB equations
 around the above Einstein gravity
configuration, with  
\begin{eqnarray}
\label{sol-gen}
v_1(\theta)=v_{10}(\theta)+\alpha_{}v_{11}(\theta)+O(\alpha)^2,~~
v_2(\theta)=v_{20}(\theta)+\alpha_{}v_{21}(\theta)+O(\alpha)^2,~~
v_3(\theta)= 1+\alpha_{}v_{31}(\theta)+O(\alpha)^2,
\end{eqnarray}
and $\bar \beta=1+\bar \beta_{1}\alpha+O(\alpha)^2$ 
(note that one can set $K=1$ without any loss of generality).
Then a straightforward computation shows that the functions $v_{i1}(\theta)$
cannot be regular at both poles of the sphere.
For example, the expression for the first order correction to the metric function
$g_{zz}$ is
  \begin{eqnarray}
  \label{solPhi}
v_{31}(\theta)=\frac{64 \pi}{3J}
\left(
-\frac{2(11+20\cos 2\theta+\cos 4\theta)}{(3+\cos 2\theta)^3}+\log\frac{2(1+\cos^2\theta)}{\sin^2\theta}
\right)
+c_1\log (\tan^2 \frac{\theta}{2}).
 \end{eqnarray}
 One can see that, for any choice of the arbitrary constant $c_1$, the function $g_{zz}$
 cannot be regular both at $\theta=0$ and $\theta=\pi$. 
 A similar result is found when considering higher orders in the expansion (\ref{sol-gen}). 
 
 Of course, regular solutions without a smooth Einstein gravity limit are not ruled out by the
 above argument.
 Therefore, we have also tried to solve non-perturbatively the set of four EGB equations 
 with suitable boundary conditions at $\theta=0,\pi$.
 However, the numerical iteration failed to converge for any finite value of $\alpha$.
 Thus we conclude that the extremal black string solutions with a regular horizon are unlikely to exist
 in EGB theory.

\section{Further remarks}  
In this work we have initiated a preliminary
investigation of the influence of the higher derivative terms in the gravity action on 
the properties of spinning black strings in $D=5$
spacetime dimensions.
Our results give numerical evidence that the well-known Kerr solution in Einstein gravity 
admits generalizations with a GB term.
Similar to the static case, these UBSs exist up to a maximal value of the
GB coupling constant $\alpha$ which depends on the event horizon radius and event horizon velocity.
Also, we have noticed that the angular velocity reduces the relative tension 
of the solutions, which approaches 
 (for fast rotating black strings) the Einstein gravity value $n=1/2$.
However, perhaps the most interesting new feature here is that
the GB term strongly affects the properties of the extremal black strings,
and seems to 
lead to some unphysical features of these configurations.

We also note an effective violation
of the weak energy condition by the UBS solutions of the EGB model.
 Here, following \cite{Kleihaus:2009dm},  we write
 the EGB equations (\ref{eqs}) as 'modified' Einstein equations,
 with an effective stress tensor that involves the
gravitational field
   \begin{eqnarray}
 R_{\mu\nu}-\frac{1}{2}Rg_{\mu\nu}=-\frac{\alpha}{4}H_{\mu\nu}=T_{\mu\nu}.
 \end{eqnarray}
 Therefore, from some point of view, the quantity  $\alpha H_t^t=-G_t^t$
 corresponds to a local
{\it `effective energy density'}.
We have found numerical evidence that this quantity  takes negative
values in some region close to the horizon.
Moreover, this region expands as the angular momentum increases and the Hawking temperature decreases.
It would be interesting to get a deeper understanding of these aspects, preferably based on
some global techniques.
 
 One should also remark that since the solutions in this work are 
 without a dependence on the extra-dimension $z$, they
 can also be
interpreted as black holes in a EGB-dilaton 
theory in four dimensions.
The action of the $D=4$ model is found by doing a reduction with respect to
the Killing vector $\partial/\partial z$ for a generic metric ansatz
\begin{eqnarray}
\label{dim-red}
ds^2=e^{-\frac{\Phi}{\sqrt{3}}}g_{\mu\nu}^{(4)}dx^{\mu}dx^\nu+e^{ \frac{2\Phi}{\sqrt{3}}}dz^2,
\end{eqnarray}
($i.e.$ with $g_{zz}=p(r,\theta)=e^{ \frac{2\Phi}{\sqrt{3}}}$)
and reads (see $e.g.$ \cite{Kobayashi:2004hq})
\begin{eqnarray}
\label{4d-S}
 I=\frac{1}{16\pi G_4}\int d^4 x \sqrt{-g^{(4)}}
\left [
R^{(4)}-\frac{1}{2}\partial_\mu \Phi \partial^\mu \Phi
+\frac{\alpha}{4} e^{\Phi}
\left(L_{GB}^{(4)}
+ 
\frac{4}{3} \partial_\mu \Phi \partial^\mu \Phi 
-\frac{1}{3\sqrt{3}}(\nabla^2 \Phi) (\partial_\mu \Phi \partial^\mu \Phi)^2
\right)
\right ].
\end{eqnarray}
The line element of the corresponding four-dimensional spinning black holes will be
\begin{eqnarray}
\label{4d-m}
ds^2_4=g_{\mu\nu}^{(4)}dx^{\mu}dx^\nu = -\hat fdt^2+\frac{\hat m}{\hat f}\left(dr^2+r^2 d\theta^2\right) 
       +\frac{\hat l}{\hat f} r^2 \sin^2\theta
          \left(d\varphi-\frac{\hat \omega}{r}dt\right)^2,
\end{eqnarray}
with $\hat f=f \sqrt{p},~\hat m=m p,~ \hat l=lp$ and $\hat \omega = \omega $ (where $f,l,m, \omega$ and $p$
are the metric functions in the five-dimensional line-element (\ref{metric})).
The properties of these solutions result  straightfordwardly from those of the
$D=5$ black strings discussed in this work. 
 
One should mention that spinning black holes of a simplified $D=4$ EGB-dilaton model
containing only the first three terms in (\ref{4d-S}) ($i.e.$
with a standard kinetic term  only for the dilaton),
and a different value of the dilaton coupling constant, 
have been discussed recently in \cite{Kleihaus:2011tg}.
As expected, they present many common features with the 
solutions in this work, 
in particular the extremal limiting configurations
being singular in both cases.
 
Similar to the $\alpha=0$ case, the generalizations  with a $U(1)$ field of the  $D=4$ spinning black holes (\ref{4d-m})
can be generated by boosting the $D=5$ UBSs in the fifth direction,
$z=\cosh \gamma~Z+\sinh \gamma \tau,~~t=\sinh \gamma~Z+\cosh \gamma\tau,$
with $\gamma$ an arbitrary parameter.
Then the dimensional reduction of a UBS configuration
along the $Z-$direction  provides new
solutions in a $D=4$ EGB-U(1)-dilaton theory, generalizing the well-known
dilatonic Kerr-Newman black holes in \cite{Gibbons:1985ac}.
In principle, based on the results in this work,
one can obtain a complete description of these solutions.
However, one should remark that due to the presence of the GB term in $D=5$,
the action of this four dimensional model has a very complicated and rather exotic
form, with non-standard terms for the dilaton and the U(1) fields (see $e.g.$
the Appendix A in Ref. \cite{Bao:2007fx}).

As avenues for further research, it would be interesting to extend the solutions in this work
by adding $n>1$ extra-dimensions ("black branes").
Based on the results in  \cite{Sahabandu:2005ma},
we expect these configurations to retain the  basic features
of the black strings studied here.
Another possible direction would be to construct spinning generalizations of the 
$D>5$ static EGB  black strings discussed in 
\cite{Brihaye:2010me} ($i.e.$ generalizations of the Myers-Perry black strings),
in which case we expect a different pattern of the solutions.

We hope to return with a systematic study of these aspects in  a future publication.

\bigskip
\noindent
{\bf\large Acknowledgements} \\
We  would like to thank Keno Eilers for useful discussions.
We gratefully acknowledge support by the DFG,
in particular, also within the DFG Research
Training Group 1620 ''Models of Gravity''.

 
 \begin{small}
 
 \end{small}

 \end{document}